\documentclass[sigconf]{acmart}
\usepackage{tabularx}
\usepackage{booktabs} % for nicer tables
\usepackage{xcolor}
\usepackage[normalem]{ulem} % keeps \emph from becoming underlined

\AtBeginDocument{%
  }

\copyrightyear{2026}
\acmYear{2026}
\setcopyright{cc}
\setcctype{by}
\acmConference[CHI '26]{Proceedings of the 2026 CHI Conference on Human Factors in Computing Systems}{April 13--17, 2026}{Barcelona, Spain}
\acmBooktitle{Proceedings of the 2026 CHI Conference on Human Factors in Computing Systems (CHI '26), April 13--17, 2026, Barcelona, Spain}
\acmPrice{}
\acmDOI{10.1145/3772318.3791571}
\acmISBN{979-8-4007-2278-3/2026/04}

 \newcommand{\new}[1]{#1}

 \newcommand{\remove}[1]{}

\newcommand{\farnazcolor}{black}

\newcommand{\colortxt}[1]{\textcolor{\farnazcolor}{#1}}

\newcommand*{\TitleParbox}[1]{\parbox[c]{1.5cm}{\raggedright #1}}%

\begin{document}

\title{Interpretive Cultures: Resonance, randomness, and negotiated meaning for AI-assisted tarot divination}

% TODO determine rules for lab/more specific affiliation
\author{Matthew Kieran Prock}
\authornote{Equal contribution}
\email{prockmtt@umich.edu}
\affiliation{%
  \institution{The University of Michigan}
  \city{Ann Arbor}
  \state{Michigan}
  \country{USA}
}

\author{Ziv Epstein}
\email{zive@mit.edu}
\authornotemark[1] 
\affiliation{%
  \institution{MIT}
  \city{Cambridge}
  \state{Massachusetts}
  \country{USA}}

\author{Hope Schroeder}
\email{hopes@mit.edu}
\affiliation{%
  \institution{MIT}
  \city{Cambridge}
  \state{Massachusetts}
  \country{USA}}

\author{Amy Smith}
\email{amyelizabethsmith01@gmail.com}
\affiliation{%
 \institution{Queen Mary University}
 \city{London}
 \country{England}}

\author{Cassandra Lee}
\email{cass_lee@media.mit.edu}
\affiliation{%
  \institution{MIT}
  \city{Cambridge}
  \state{Massachusetts}
  \country{USA}}

\author{Vana Goblot}
\email{v.goblot@gold.ac.uk}
\affiliation{%
  \institution{Goldsmiths, University of London}
  \city{London}
  \country{England}}

\author{Farnaz Jahanbakhsh}
\email{farnaz@umich.edu}
\affiliation{%
  \institution{The University of Michigan}
  \city{Ann Arbor}
  \state{Michigan}
  \country{USA}}

\renewcommand{\shortauthors}{Prock et al.}

\begin{abstract}
While generative AI tools are increasingly adopted for creative and analytical tasks, their role in interpretive practices, where meaning is subjective, plural, and non-causal, remains poorly understood. This paper examines AI-assisted tarot reading, a divinatory practice in which users pose a query, draw cards through a randomized process, and ask AI systems to interpret the resulting symbols. Drawing on interviews with tarot practitioners and Hartmut Rosa's Theory of Resonance, we investigate how users seek, negotiate, and evaluate resonant interpretations in a context where no causal relationship exists between the query and the data being interpreted. We identify distinct ways practitioners incorporate AI into their interpretive workflows, including using AI to navigate uncertainty and self-doubt, explore alternative perspectives, and streamline or extend existing divinatory practices. Based on these findings, we offer design recommendations for AI systems that support interpretive meaning-making without collapsing ambiguity or foreclosing user agency.
 \end{abstract}



\begin{teaserfigure}
 \centering
  \includegraphics[width=0.85\linewidth]{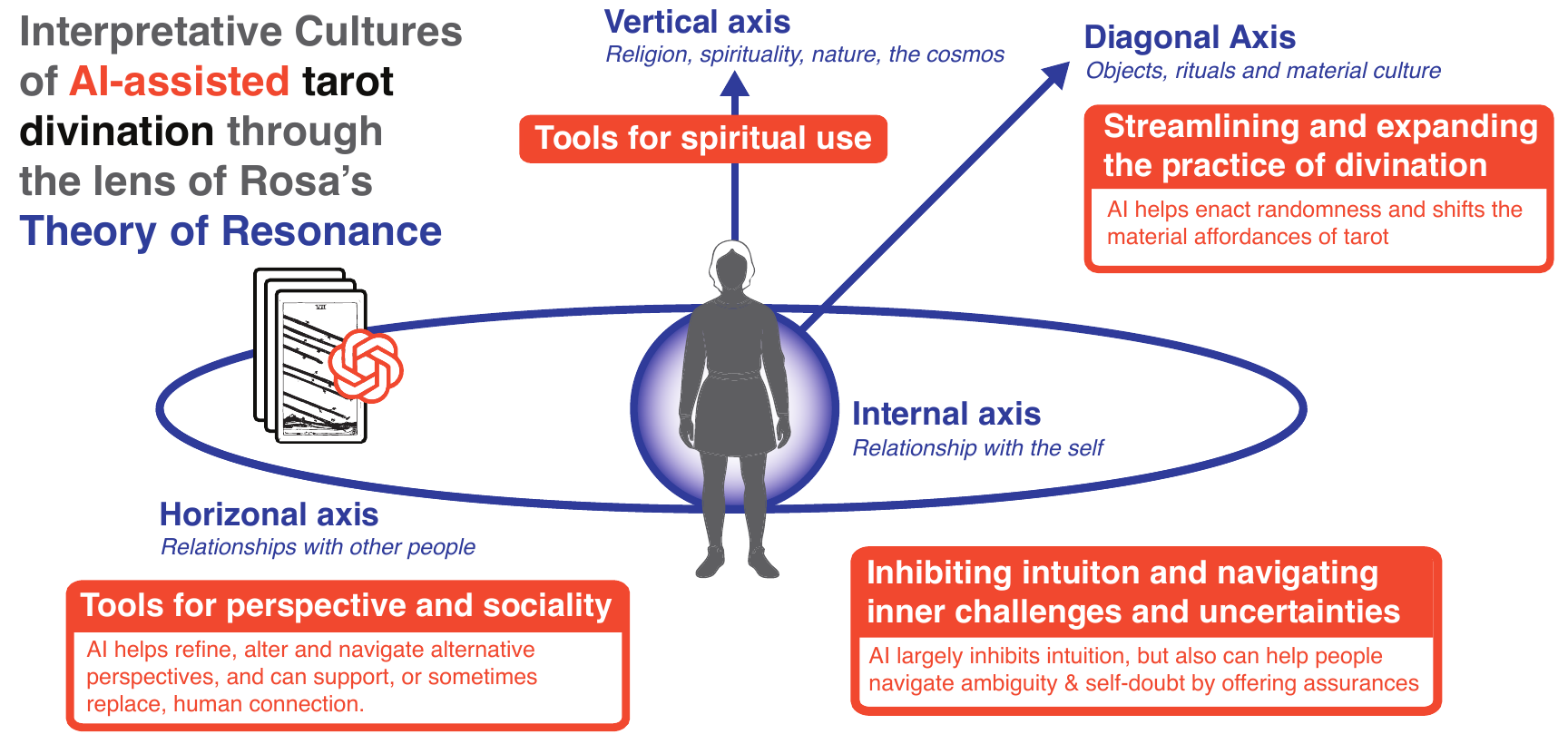}
  \caption{Interpretative cultures of AI-assisted tarot divination through the lens of Rosa's Resonance Theory. Blue: The four axes of Resonance Theory (see Section 3). Orange: an overview of the results of the interview study (see Section 5).}
  \Description{Interpretative cultures of AI-assisted tarot divination through the lens of Rosa's Resonance Theory. The four axes of Resonance Theory. Engaging with tarot through AI, we theorize, can activate on these axes. We report on how the ritual practice of randomness in general, and of tarot in particular enacted by AI, are core to fostering resonance (diagonal axis). AI use in tarot helps along the horizontal axis by refining, altering, and navigating a variety of alternative perspectives. As this axis also concerns sociality, AI can support or sometimes replace human connection when used for tarot (horizontal axis). With respect to the internal axis of resonance, AI largely inhibits intuition, but it can also help people navigate ambiguity or self-doubt by offering assurances.}
  \label{fig:resonance-axes}
\end{teaserfigure}

\maketitle

\section{Introduction}
As large language models (LLMs) and other generative AI systems have increased in capabilities, we have seen the mass adoption of AI tools in a wide variety of creative and subjective tasks and increased scholarly attention to their downstream effects\cite{epstein2023art}. But for a class of tasks that we refer to as \textit{interpretive work}, i.e. the interpretation of data relative to a given query, the promises and perils of AI interaction remain largely understudied and poorly understood.

In this paper, we are interested in exploring the use of AI in interpretative tasks \new{where meaning is generated in the absence of any causal link between a query and the data used to answer it.}\remove{that exhibit \textit{polysemy}---the possibility of multiple, perhaps divergent meanings. One domain where interpretative ambiguity is central} \new{This dynamic is most clearly seen in} \textit{divination}, the ritual practice of gaining insight into a query by interpreting patterns drawn from \remove{randomness. In particular, divination is an example of \textit{synchronicity}, a concept devised by Carl Jung in collaboration with Physics Nobel Laureate Wolfgang Pauli to describe events that uniquely coincide but lack causal connection. One example of synchronicity is divination with tarot cards, in the sense that the cards that are drawn from the shuffled deck (i.e., the data) and the nature of the user's query are necessarily independent from each other by construction \cite{semetsky2010folds}. A central feature of divination is the use of} random procedures, such as shuffling and drawing cards or tossing coins. From a statistical perspective, these procedures deliberately sever any causal link between the querent’s question and the outcome, producing data that is statistically independent by design and potentially infinitely generative. In practice however, the ritual act of shuffling/drawing cards or flipping coins is rarely experienced as neutral: people often contemplate their question or focus their intention while handling the cards. The combination of ritualized randomness and subjective intentionality creates a paradoxical space in which outcomes can feel both unpredictable and deeply personal. 

This non-causal structure distinguishes divination from other interpretative tasks---such as qualitative coding, clinical evaluation, biblical hermeneutics, or psychoanalysis---where the data bears a causal connection to the query. In those cases, meaning can be derived through deduction and inference. While subjectivity is always present, there is some expectation of intersubjective agreement (that of course varies \new{in degree} from domain to domain), where attentive interpreters often converge on overlapping patterns. In divination, while interpretive traditions and shared symbolism can support a degree of common understanding, the causal structure between the query and the data is intentionally cut \cite{dove1993uncertainty}. \new{Despite its cultural visibility, divination---and especially its contemporary digital forms---remains largely absent from HCI scholarship. Existing work on meaning-making with AI has focused on domains with stable ground truth or shared interpretive norms, leaving open how users use AI systems in contexts defined by ambiguity, symbolism, and intentional randomness.}  

\remove{Divination is an understudied interpretive domain within HCI. What was once regarded as a countercultural niche practice that practitioners did in secret has become a vanguard cultural practice in the 21st century, popularized by social media and digital cultures.} A largely popular form of divination in the West is tarot, a set of playing cards each with a name and evocative imagery, which practitioners consult at random in relation to a particular query. 
\new{Our team's familiarity, at varying levels, with contemporary tarot subcultures has made visible to us a growing pattern of practitioners relying on AI tools such as ChatGPT to help interpret cards drawn in personal readings.
 While far from a universal or uncontested practice, these reports signal an emerging shift that warrants examination as part of evolving human–AI meaning-making practices.}
\remove{This project was born out of the authors' observation that some individuals are now reporting using an AI-based tool such as ChatGPT to help them interpret the tarot cards they pulled in a personal reading.} What may be surprising about the use of AI for interpreting tarot is that AI cannot rely on the cards or the querent’s question alone, since by design \remove{it is a random process and therefore }there is no underlying correlation\new{al or causal structure} to uncover (though we acknowledge that some \remove{cosmologies}\new{belief systems} around tarot do not believe the outcomes to be random). Instead, the AI system “hallucinates” meaning by drawing on the wide body of discourse it has absorbed about tarot, human experience, and beyond, steered by the user’s written query or instruction to the AI model as well as potentially stored memory of the user’s past interactions. Human readers likewise bring in more than just the card and the query, but they typically draw on specific traditions, personal intuition and experience, and their \remove{chosen cosmology}\new{belief system}. This raises the question of how people who use AI in this task come to find its interpretations resonant---whether they accept them as meaningful out of the box, or whether resonance with a particular interpretation emerges through a process of negotiating and reshaping the AI’s output. 
% We also investigate if people are using it in similar ways as tools that have already existed (websites and people) or in new ways (on demand dialogue with a totalized ``Other'').

In this paper, we explore how the production of meaning in the context of divination is shaped by the introduction of AI tools such as ChatGPT. What role does AI play in the divination process? How do users negotiate and co-construct meaning with AI systems as they interpret tarot spreads in relation to their query? What forms of meaning is produced through this ritual AI-assisted  \new{practice of randomness}? 

We approach these questions through interviews with tarot practitioners who use AI to identify how users negotiate and make meaning with AI systems and what patterns of meaning-making exist in these rituals \new{that are based on} randomness. We find that our participants used AI during the interpretive workflow in a variety of ways, with AI empowering users to navigate inner challenges or uncertainties, providing alternative perspectives that deepen their interpretations, and also expanding their existing practice of divination.

\new{To articulate these processes, we draw on Hartmut Rosa's sociological Theory of Resonance, which provides a framework for understanding how people seek, experience, and evaluate meaningful contact with the world. Building on this framework, we trace tarot practitioners’ engagements with AI along four axes of resonance---the internal, horizontal, diagonal, and vertical---to describe how AI supports self-reflection, offers alternative perspectives, streamlines the enactment of divinatory ritual, and situates meanings within spiritual or cosmological orientations.}
\remove{To do so, we first draw on Hartmut Rosa's sociological Theory of Resonance as a framework for uncovering the axes of meaning-making involved in interpretive labor. We connect Resonance Theory to discussions of randomness as a way to bridge resonance and the use of generative AI in the domain of divination.} 

% How do their cosmologies and sense-making practices extend to the use of AI tools? 

% We also found that participants’ beliefs about the cosmology of tarot shaped not only how they incorporated AI into their practice, but also how they managed interpretations that did not resonate and how they conceptualized AI's role in divination. 
We observe that these patterns of meaning production form a newly emerging \textit{interpretive culture}. Finally, we conclude with design recommendations for AI systems used in the interpretative process, particularly those for interpretative tasks for which no causal connection exists between the semiotic signal and the query. \new{This study offers an exploratory probe for concept-building toward understanding the  dynamics that shape how people work with AI in interpretive domains.} To summarize, our contributions are:
\begin{enumerate}
    \item A novel interdisciplinary approach to theorizing interpretive meaning-making that draws on cultural studies and Rosa's Theory of Resonance
\item Empirical understanding of how users construct meaning with AI systems, identifying a newly emerging interpretative culture
\item A series of design recommendations for the design of AI systems supporting symbolic and non-causal interpretive practices, inspired by insights from this interpretative culture
\end{enumerate}
\begin{figure*}[h]
    \centering
    \includegraphics[width=0.99\linewidth]{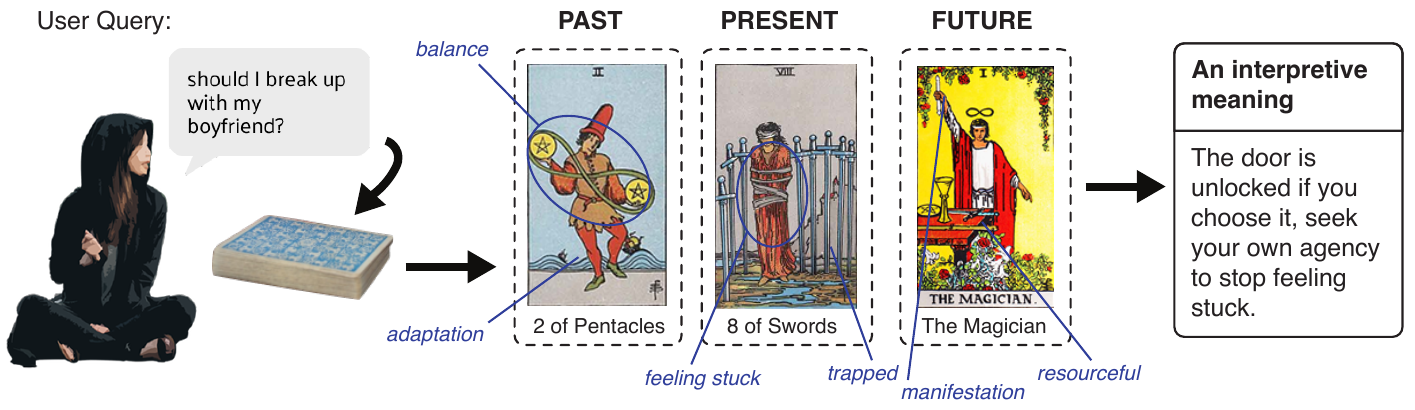}
    \caption{An overview of the standard tarot reading process}
    \Description{This figure shows a 3 card spread for a user who uses tarot to answer the question "should I break up with my boyfriend?". This is a 3 card spread. The card representing Past is 2 of Pentacles which can mean balance or adaptation. The card representing Present is 8 of Swords which can mean feeling stuck or trapped. The card representing Future is the Magician, which can mean manifestation or resourceful. A possible interpretation of this draw is: The door is unlocked if you choose it, seek your agency to stop feeling stuck.}
    \label{fig:primer}
\end{figure*}

\section{Related Work}
\subsection{Tarot as a cultural practice}
Tarot is a deck of 78 ornate cards, distinctive in that it includes 22 trump cards (``Major Arcana'') alongside 56 suit cards (court and pip cards), collectively known as the Minor Arcana. It has been used for a wide range of practices, including divination, self‑reflection, and creative world‑making \cite{au2025mirror, dummett1980game, farley2009cultural} (see Figure \ref{fig:primer} for an overview). The deck’s origins can be traced back to the Italian Renaissance, when tarot functioned primarily as a card game \cite{dummett1980game, moakley1966tarot, farley2009cultural}. From there, it circulated across Europe and, from the eighteenth century onward, became increasingly entangled with esoteric and divinatory practices, frequently reinterpreted and infused with mystical symbolism \cite{dummett1980game, moakley1966tarot, farley2009cultural}. In a typical reading, the practitioner draws cards into a structured spread---ranging from a single-card pull to more elaborate layouts such as the Celtic Cross---where each position provides an interpretive frame (e.g., past influence, present concern, future trajectory). Interpretation then involves drawing on the card’s established symbolic associations---for example, a card like the Magician, which can signal manifestation, resourcefulness, or creativity---supplemented by its imagery, and considering how these interact with the querent's circumstances (see Figure \ref{fig:primer}).

Since the twentieth century, the proliferation of tarot deck production and consumption has contributed to tarot’s consolidation as a growing cultural practice, as decks and readings have become widely available and are now used for many purposes in a wide range of settings, including not only fairs and shops but also online marketplaces and social media platforms such as Etsy, YouTube, TikTok, where audiences of varying size are invited to ``take the interpretation as it resonates,'' a recurrent catchphrase among contemporary online tarot readers \cite{cartwright2020tarot, wen2023state, pew2025newage}.

\subsection{The rise of mainstream divination}
Scholars have described contemporary society as post-secular, a term that points to the persistence and renewed visibility of spiritual and religious practices in contexts once assumed to be increasingly secular \cite{taylor2007secular, lyon2010being, habermas2008notes}, a period in which the once-rigid boundaries between secularism---characterized by the rise of science and technology---and spiritual beliefs are waning. Rather than encouraging divisions between faith and reason, Charles Taylor \cite{taylor2007secular} and Jurgen Habermas \cite{habermas2006religion} observed that the interaction of spiritual and secular worldviews now shape cultural norms and identities. These new ``social imaginaries''~\cite{taylor2007secular} of individualized positionalities, such as “spiritual but not religious” identities, are signaling toward a renewal in uptake of activities such as tarot or astrology, offering a plethora of practices of divination which are now entwined with practices of therapy and self-improvement \cite{auyeung2025mirror}. Indeed, according to a 2025 Pew study, 30\% of Americans now use astrology, tarot, or fortune tellers at least once a year \cite{pew2025_astrology}. 

\subsubsection{Divination within HCI} From an HCI perspective, this rise of mainstream focus positions divination as a site of everyday interaction with symbolic systems, where people use artifacts and practices to construct meaning. Past work in HCI has explored several aspects of divination. In the context of tarot, \citet{michelson2024worlding} examine currents of deck design as they connect with larger visions of digital futures. They surface themes of personal/collective growth, carework, and pathways to alternative ways of knowing as tools to expand the scope and nature of designerly knowledge production. Relatedly, the Tarot Cards of Tech were introduced as a speculative design deck that adapts tarot's structure to help designers reflect on assumptions, values, and potential societal impacts of new technologies~\cite{tarotcardsoftech}. \citet{lustig2022explainability} create a machine learning (ML)-generated deck of tarot cards to shift conversations on AI explainability from a focus on transparency toward ineffability---the sensory engagement of aesthetic experience \cite{boehner2008interfaces}. Beyond tarot, HCI researchers have developed several devices for divination \cite{akmal2020divination}. \citet{benabdallah2020sybil} develop a divinatory home device that delivers AI-generated prophecies based on the participants' breathing patterns. \citet{altarriba2020mesmer} repurpose the Ouija board to develop a tangible conversation tool for playful design building on the Otherworld Framework \cite{borutecene2019otherworld}. Elizabeth Buie's work in HCI has explored techo-spirituality \cite{buie2013spirituality} and the design of ``transcendent user interfaces'' \cite{buie2018exploring}.

\subsection{Interpretive labor}
In her ethnographic study of tarot readers, Karen Gregory examines how tarot readers' spiritual practice relies on interpretative labor \cite{gregory2012negotiating, gregory2019pushed}. She argues that interpretation of tarot cards is not simply rote symbolic translation---it is a demanding, ongoing negotiation of meaning, emotion, and relational connection. Gregory details how tarot practitioners must work with a complex range of ambiguous symbols, with each reading engaging them with unpredictable contexts and specific circumstances, within which they are forced to negotiate \remove{own embodied insight} \new{their intuitive understanding} with traditional meanings of cards and specific queries.

The process that Gregory outlines evokes Mikhail Bakhtin’s concept of dialogic imagination \cite{holquist2003dialogism}, which conceptually bridges tarot readers, spiritual workers and now, querents consulting AI in their readings. The meaning-making processes that arise out of the consultation with cards takes place through a form of dialogue which incorporates active exchange of perspectives and contexts. This interpretative process involves what Bakhtin calls ‘heteroglossia’--- competing interpretations, \remove{intensions} \new{meanings}, and values that inform interpretation which involve both affective and cognitive labor.

\subsection{Interpretive cultures}
Randomness structures the divinatory practice itself, which in turn gives rise to a social and cultural context in which interpretation takes place, mediated and extended by modern technology. In her book \textit{Life on Screen} \cite{turkle1995life}, Sherry Turkle argues that our interactions with technology are always informed and shaped by meanings and beliefs we attribute to them. In this view, technology enables a new social and cultural sensibility, and in the case of tarot and divination, these sensibilities are now extended and mediated by generative AI which, according to our research, is mainly used for the purposes of interpretative labor. This is because tarot has been a ‘meta-semiotic system’ \cite{aphek1990semiotics}. This means that, following Aphek and Tobin’s account, tarot is a system  in which signs carry a web of  visual, textual, and contextual meanings that are also shaped by ceremony (ritual) and specific connection with the querant but generate further interpretive possibilities. Situating generative AI within this framing highlights how its outputs may be taken up as additional layers of symbolic material, extending the interpretive labor that has long characterized divinatory practice.

This divinatory tool has traditionally been seen as possessing secret or hidden knowledge that implies esoteric insight. The most commonly used systems in circulation are modeled on the Rider Waite Smith tarot deck \cite{decker1996wicked}. Marseilles Tarot (MT), first produced in 17th century France, remains as one of the dominant tarot semiotic systems - while the more recent Thoth deck (TT), first published in 1969, is also gaining traction and popularity. With thousands of different decks currently in circulation, some of which use the existing layouts and interpretive frameworks of RWS, MT or TT, we have to take into account not only the recontextualisation of images and symbols, or even the invention of new systems that include `oracle decks', but also individual autonomy in the ways that preferred readings are organized. All this has contributed to changing expectations of what `tarot reading' is - from prescribed meanings for each card, to meanings that develop in relation to new deck designs, but also through valuing the creative interpretative skills of readers that accept multi-layered meanings that sit anywhere on the spectrum between prescribed (using tarot guides) to being free-form; from being generalized, to being thematically organized. The exponential popularity of tarot readings on social media has further normalized this liminal knowledge as \remove{exoteric} \new{mainstream}, not only esoteric, culture \cite{auger2016cartomancy, fu2022techno_cultural_domestication} contributing to formations of \emph{interpretative cultures} around popular tarot influencers, who are developing their own styles and cultures of interpretation, often using a combination of different decks and divination practices and relying on intuitive, non-traditional and storytelling style aspects of readings. From an HCI perspective, this illustrates how interaction design can account for user autonomy in constructing meaning, where artifacts provide resources rather than prescribing fixed interpretations. \remove{So there is a case for the meanings not solely being encoded ‘in the cards’; the triumvirate of the querent, the reader, cards (with the latter two being replaced by AI), is driven by social and cultural context out of which these intimate consultations take place. Guidebooks offer suggestions, but do not prescribe specific advice. The cards offer both structured rules for reading, or for the “sensitive” an opportunity to explore “inclinations and knowings” \cite{falkner2020tarot}. The card meanings are constructed and produced in a plurality of contexts even more so with availability of general(ized) readings, encouraging both recipients and readers possess the capacity of relative autonomy in their interpretation, clearly illustrating Stuart Hall’s production of polysemic meaning making processes; there are many ‘preferred’ readings encoded in many tarot decks, and equally the reading positions is expected to lean towards negotiated or sometimes even oppositional meanings, with some readers or recipients of readings resisting them. The case of tarot highlights how interpretive practices may be shaped not only by symbols and traditions but also by technological mediation.} To situate this within a broader frame, it is useful to consider how AI is being taken up in other forms of subjective labor.

\subsection{AI and subjective labor}
As large language models (LLMs) and other generative AI systems have increased in their apparent capabilities, we have seen the mass adoption of AI tools in a wide variety of subjective tasks. Some tasks, like the use of AI for artistic production, have been met with widespread critique and concern, citing problems of theft of training data \cite{gero2025creative}, shifts in responsibility \cite{epstein2020gets}, and homogenization of culture \cite{anderson2024homogenization}.

\remove{One branch of work related to has demonstrated the effects of AI when it is employed, particularly in creative applications.} 
\citet{gero2025creative} uncover attitudes that creative writers hold. Past work has shown that the use of generative AI can increase the number of outputs, but reduce the diversity of those outputs in the context of short stories \cite{doshi2024enhances_creativity}, idea generation \cite{anderson2024homogenization} and online art websites \cite{zhou2024human_creativity_art}. Underlying these concerns is philosophical tensions about the ways of knowing AI systems and their use implicitly inherent. Haghighi et al \cite{haghighi2025ontologies} make a case for ontology-aware design, and highlight four orientations for ontological engagement: pluralism (in response to the universalism), groundedness (in response to abstraction), liveliness (in response to fixity), and enactment (in response to dilution). 

Other knowledge work tasks, like programming, have been met with more optimism about AI's ability to carry out the labor or support humans in their cognitive labor. However, though AI interaction has been shown to increase efficiency and reduce errors \cite{peng2023impact}, results have so far been mixed, and AI may counterintuitively actually slow developers down \cite{becker2025measuring}. \remove{These unintended effects are also being borne out in other domains.}

A growing body of work explores how LLMs affect the meaning-making tasks of qualitative research and subjective annotation, some through human-computer user interfaces that bring AI suggestions to an analytical or interpretive task. Growing evidence from work providing AI assistance in subjective annotation tasks shows that while AI suggestions and assistance in these tasks can improve measures of user satisfaction, users strongly uptake AI suggestions, homogenizing their findings, and also actually slow down, likely as a result of parsing additional information in the form of a suggestion from AI \cite{overney2024sensemate, schroeder_just_2025}. \citet{schroeder2025large} explore tensions and intentions when using LLMs in qualitative research through interviews with researchers across academic communities. They find that while qualitative researchers are experimenting with LLMs and hope they can be useful in tedious tasks, many worry that the imposition of external perspective from an LLM not only \remove{risks their methodologies} \new{may undermine the methodological rigor of their approaches} (such as grounded theory) but also risks misinterpretation or harm when interpreting data, particularly from vulnerable or otherwise marginalized groups. 

In parallel, growing efforts to recognize the subjective nature of interpretive annotation have made their way into subfields of AI development, one of which is perspectivist NLP. Whereas in traditional AI development, annotator disagreement on subjective tasks, like hate speech detection, was \new{considered as noise to be} minimized, perspectivist NLP acknowledges that varied, and even disagreeing, perspectives when applying subjective, interpretive labels to data can be important signal about legitimate but differing interpretations from individuals involved in the interpretive work \cite{sap2021annotators, frenda2025perspectivist, cabitza2023toward}. Some emerging approaches to AI development thus build in probabilities for multiple possible interpretations to be predicted from a single system.

\section{Situating randomness and divination through the lens of Rosa’s Theory of Resonance 
}
Scientific knowledge systems hold materialism as a fundamental assumption: that observation and experimentation are necessary and sufficient to develop explanatory and mechanistic theories of the world. Randomness is treated as an impediment---something to be minimized through increased control in experimental design. Divination, on the other hand, is a non-scientific method of knowledge acquisition rooted in an alternative theory of randomness. This alternative \remove{cosmology} \new{belief system} holds that ``the only constant in the universe is change'', and that the randomness we observe, however chaotic or coincidental, is a veil that hides \emph{correspondencies}. These correspondencies are taken to be meaningful connections waiting to be discovered. \remove{For C. G. Jung, synchronicity is the process of meaning making where correspondence draws meaning out of randomness~\cite{pennick1989_gamesofgods, faivre2000theosophy}.}

In this context, randomness in divination plays a dual role: it removes deterministic inference while also supporting the authority of the interpretation by allowing meaningful patterns or correspondencies to emerge between the random outcome and the query. The apparent arbitrariness of whatever meaning emerges through interpretation is not experienced as fully self-generated. Instead, it is attributed to chance, the world, or a transcendent force. For many practitioners, this separation from conscious choice is precisely what makes the outcome believable or resonant: the signs feel authorized by a process outside their direct control, allowing the interpretation to carry weight without requiring proof of causal linkage.  

\citet{dove1993uncertainty} describes how the Kantu’ of Borneo use bird augury rituals that deliberately minimize correlation between human intention and the observed signs. By ensuring that the signs are produced outside of the individual’s control (through the contingent behavior of birds in nature) the outcome is positioned as impartial, since no human could have engineered it. This perceived impartiality, in turn, lends authority to the interpretation as the signs can be treated as carrying a message that exceeds the individual’s own will or bias. The specific way this authority is understood depends on the belief system at play. Within some belief systems, the absence of correlation is taken as evidence that a transcendent force is speaking through the signs; within others, it is seen as opening a space for self-reflection, where the random event functions like a mirror for the querent’s concerns. Whichever the cosmology, both the absence and the presence of correlation in this instance are interpreted as meaningful, or as we want to posit here, resonant connection to the world. 

For Hartmut Rosa, a resonant relationship with the world presupposes ‘a kind of mutual, rhythmic oscillation’  through ‘a different way of relating to the world’~\cite{rosa2019resonance}. Resonance is a way of attuning to randomness. 
% It serves as a counterforce to the imbalance brought by the pressures of social acceleration in which modernity is ‘out of tune’ and conditions ‘mute’ relations and  the ‘moments of extreme thrownness in which the world confronts us as hostile and cold’.
It is a powerful response to the pressures of modern life, which often leave people feeling disconnected, rushed, or emotionally flat. In these conditions, our relationship to the world can feel ``mute''---as if the world no longer speaks back---and we may find ourselves in moments of what he calls ``extreme thrownness,'' when life feels cold, indifferent, and beyond our control~\cite{rosa2019resonance}.
Resonance offers the vibrant mode of relating to the world, which operates through distinct social spheres of activity and experience. Rosa outlines four ‘axes of resonance’ or directions through which resonance can occur (see Figure 1, orange):
\begin{enumerate}
 \item \textbf{Internal resonance} (or self-resonance)  provides an account of understanding our own emotions and desires, meaningful connection with the self through introspection, therapy, and divination.
    \item \textbf{Horizontal resonance} features in relationship to other human beings.
    \item \textbf{Diagonal resonance} describes the meaningful engagement with artifacts, or specific activities and rituals, such as working with tarot decks or AI for tarot readings---where reciprocal transformation occurs between practitioner and object.
    \item \textbf{Vertical resonance} is a response to ‘the loss of metaphysical axes of resonance in the sense of a cosmological or theological resonant order’ and is characterized by the reestablishment of the relationship with spirituality, nature or art~\cite{rosa2019resonance}.
\end{enumerate}

   All of these axes are related and connected. According to Rosa, they are a ``living network'' that can \new{support, shape, and sometimes conflict with each other.}

Rosa’s theory of Resonance has proven a productive lens through which to study domains such as education \cite{prouteau2022critique} and nursing~\cite{lopez2023social}. Here, we extend its applicability to divination and randomness. We theorize that drawing on Rosa’s Theory of Resonance as applied to divination and randomness can also provide insight into new ways of designing interactions with AI systems. In particular, encounters with AI are encounters with a different kind of alterity, but an alterity nonetheless. AI systems are also systems of randomness, or ``stochastic parrots'', albeit recent engineering efforts to increase technological determinism in the service of convergent values have come at the cost of randomness, inducing concerns of algorithmic monoculture and homogenization \cite{ugander2024art}. In the face of the potential crisis of meaning and alienation that LLMs induce, Rosa’s Theory of Resonance gives us a framework for designing AI interactions that could resonate across \new{the}\remove{all four} axes, which we unpack in Section 6.1.

\section{Method}
In order to investigate how AI is applied to \remove{the polysemic interpretative task of} divination, we conducted \remove{11}\new{12} semi-structured interviews with \new{practicing tarot readers who use AI as part of their personal divination work.}\remove{individuals who currently apply AI to their personal tarot and divination practices.}\remove{We analyzed the interviews using thematic analysis \cite{braun2022thematic} and inductive coding methods.} This study was approved by our Institutional Review Board (IRB).

Participants were recruited using purposive sampling techniques \cite{bernard2013sampling} through public, semi-private, and personal networks in order to \remove{reach saturation in the interview data.}\new{to ensure a sufficiently wide participant sampling pool to identify thematic stability.} We shared a digital flyer within Reddit and Discord groups which were selected by keywords, including \textit{tarot}, \textit{AI}, \textit{divination}, \textit{tech}, \textit{artificial}, \textit{intelligence}, \textit{lenormand}, and \textit{witch}. We also individually reached out to users on Reddit, Instagram, and TikTok who had previously shared posts describing their application of AI to their tarot or divination practices. In addition, we posted our physical flyer in a local tarot shop. Participants had to be at least 18 years of age, \remove{ not located in the UK or EU countries during the interview in order to ensure our research practices aligned with university guidelines regarding current GDPR constraints,} practice tarot and/or divination, and apply AI to their personal tarot or divination practices in order to be eligible. \new{Additionally, because our study involved cross-institution data sharing originating from a US institution, our approved IRB protocol did not satisfy GDPR's requirements for processing and transferring personal data from those regions. As a result, we were not permitted to enroll EU or UK participants and therefore excluded participants from the UK and EU for this study.} Within the recruitment survey, we required that participants answer three short-form questions which asked them to recall their experience surrounding a recent tarot reading, describe how they use or have used AI to help interpret a tarot spread, and to briefly explain their interest in participating in the study.  We only reached out to those who answered all of the questions.

We collected demographic information about the participants pertaining to their age, race/ethnicity, gender, level of education, and current country. \new{We have also included how long each participant has been practicing tarot, as well as how each participant self-described their current skill level practicing tarot; three participants considered themselves beginner tarot readers (P4, P8, P12), six as intermediate readers (P2, P3, P6, P7, P10, P11), and the rest as expert readers (P1, P5, P9).}

\TitleParbox{}
\begin{table*}[ht]
    \centering
    \begin{tabularx}{\textwidth}{c l l l l l l}
    \toprule
    \TitleParbox{\textbf{Participant}} & \TitleParbox{\textbf{Age Range}} & 
    \TitleParbox{\textbf{Gender Identity}} & \TitleParbox{\textbf{Education}} & \TitleParbox{\textbf{Length of Tarot Practice}} & \colortxt{\TitleParbox{\textbf{Tarot Skill Level}}} & \TitleParbox{\textbf{Country}} \\
    \midrule
    1  & 45--54 & Male & Master’s & >30 years & \colortxt{ Expert} & United States \\ 
    2  & 18--24 & Nonbinary & Bachelor’s  & 3 years & \colortxt{ Intermediate} & United States \\
    3  & 35--44 & Female  & Professional Degree &  2 years & \colortxt{Intermediate} & Canada \\
    4  & 25--34 & Female & Bachelor’s  & 11 years & \colortxt{ Beginner} & Canada \\
    5  & 35--44 & Female  & Bachelor’s   & 12 years & \colortxt{ Expert} & Mexico \\
    6  & 25--34 & Male  & Some college  & 14 years & \colortxt{Intermediate} & Mexico \\
    7  & 18--24 & Male & Some college  & 2 years & \colortxt{ Intermediate} & United States \\
    8  & 18--24 & Male  & Some college  & 1.25 years & \colortxt{ Beginner } & United States \\
    9  & 35--44 & Nonbinary & Associate’s  & 20 years & \colortxt{ Expert } & 
    United States \\
    10 & 35--44 & Male & Bachelor’s & 2 years & \colortxt{ Intermediate } & United States \\
    11 & 25--34 & Female & Master’s & >10 years & \colortxt{ Intermediate } & United States \\
    \colortxt{ 12 }& \colortxt{25--34} & \colortxt{Female} & \colortxt{Master’s} & \colortxt{2 years} & \colortxt{Beginner} &  \colortxt{ Canada }\\
    \bottomrule
    \end{tabularx}
    \caption{Participants' demographics (age range, gender identity, education level, length of tarot practice, country)}
    \Description{Participants' demographics (age range, gender identity, education level, length of tarot practice, country). 25\%\ of participants were 18-24, 33\%\ were 25-34, 33\%\ were 35-44, and the remaining 9\%\ was 45-54. 42\%\ of participants identified as male, 42\%\ as female, and 16\%\ as nonbinary. Of the degrees participants obtained, 25\%\ were Master's degrees, 33\%\ were Bachelor's degrees, 17\%\ were associate's or professional degrees, and 25\%\ of participants had completed some college. Half of the participants had practiced tarot for under 10 years, 42\%\ for 10-20 years, and one participant for over 30 years. By tarot skill level, 25\%\ considered themselves beginners, 25\%\ intermediate, and the rest as experts. 25\%\ of participants were located in Canada, 17\%\ in Mexico, and the rest were located in the United States.}
\end{table*}

 Our interview questions were designed to understand each participant’s typical tarot practices, particularly how they incorporate AI, and their perspectives on how AI has impacted their personal practices. Interviews took place over Zoom; audio and video files were saved automatically after the interviews were concluded. These recordings were used in order to clean the transcript of the interview automatically generated by Zoom, and were deleted immediately afterward. We began each interview with questions surrounding each participant’s background with tarot, including how long they have practiced tarot or divination and their skill or comfort level, and then focused on the processes they typically applied to interpret a reading. After gaining an overview of their tarot practices, we honed in on their applications of AI to tarot, in order to develop an understanding of what aspects of their practice were scaffolded by the AI tools that they used. \new{Interviews on average lasted approximately 60 minutes.} We analyzed the interview data through inductive thematic analysis. Throughout the data collection process we iteratively coded participant interviews, meeting several times to discuss the themes, and eventually reaching the point where key themes emerged (see Section 4.1 for details). Each transcript was reviewed more than once, in order to ensure coding accuracy. 

 % \new{
 % We completed a second wave of data collection and the main identified themes remained stable indicating theoretical saturation \cite{glaser2017discovery}. 
 % }

Throughout the recruitment process, our team encountered a few challenges, mostly regarding the polarity of generative AI within the tarot community. When reaching out to the online spaces described above to request permission to advertise the screening survey, we were often denied advertising the survey solely because our study was related to AI---despite the fact users within those communities shared posts about applying AI to tarot or divination. This was sometimes because forum moderators anticipated that posting about the study would result in controversy and argument. Some moderators were personally against the use of AI for tarot, or in general, which led to them denying us approval to advertise the screening survey in their community. We speculate that many users who applied AI to their tarot practice chose not to post about those experiences publicly because of this same stigma, after seeing other members of their community receive mixed reactions when sharing their practices. Even without considering the divided opinions on AI within the tarot community, we recognize that tarot and divination carry their own stigmas, and are often deeply personal practices. Users may have felt reserved about practicing tarot and were hesitant about sharing their experiences online, to avoid receiving judgment from people who misunderstand or devalue the practice. Additionally, users who turn to tarot or divination for personal questions or struggles may have also felt unwilling to disclose their practices online to protect their privacy. These two factors may have also contributed to our difficulty recruiting a large number of participants. 

\subsection{Thematic Analysis}

\new{Our analytic process followed an iterative, multi-stage thematic analysis informed by grounded theory techniques \cite{glaser2017discovery}. One member of the research team conducted all interviews in the study. This researcher then manually checked and cleaned the Zoom-generated interview transcripts. Following each interview, this same researcher conducted initial open coding of the interview transcript and generated early analytic memos and provisional codes that captured emergent concepts. The author used a combination of Atlas.ti and anonymized Google spreadsheets to track the application of the first draft of codes, which served as a foundation for the next stage of analysis. Next, multiple members of the team engaged in iterative discussion about the definitions of and applications of the codes, driven by conventional methods of ``constant comparison'' across the existing data, emergent themes, and theory recommended by conventional grounded theory methods \cite{glaser2017discovery, strauss1994grounded}. This constant comparison and refinement led to axial and selective coding \cite{strauss1990basics}, which continued through the collection of the 11th interview.}

\new{Once a stable set of high-level codes had been developed and discussed with co-authors, we synthesized the categories into a set of overarching themes, attending closely to how participants described their interpretive workflows, their interactions with AI, and the forms of meaning that emerged. In this second round of synthesis and discussion, aspects of Rosa's Theory of Resonance began to surface in the data, providing a conceptual lens that aligned with participants' descriptions. Thus, our initial, inductively derived set of codes were enriched with relevant theory through Rosa, leading to an approach to code definition that blended initial codes with theoretically-informed thematic codes used in the rest of the analysis. We finalized codes related to the four axes of resonance, which are reflected in Section 5.2.}

% We also included a code capturing broader sentiments around AI that were annotated with additional memos by the authors to provide context on their application, which is reflected in section 5.1.1}

\new{Once this iterative process of defining and applying codes had reached stability, the authors followed a process of interpretative coding~\cite{mcdonald2019reliability}, ensuring the reliable re-application of codes through consensus-driven discussion over the first 11 participant interviews. First, the 11 interviews were divided roughly evenly across four authors who engaged in thematic coding using code definitions finalized via consensus. All authors noted any instances of uncertainty with respect to the application of codes to the data, and all discrepancies were resolved via consensus discussion after independent coding. In addition to this, the two co-first authors independently double-coded their own assigned interviews and a sample of interviews drawn from the other two authors' subsets. This additional subset served as a check on the authors' application of codes. The two co-first authors compared their application of codes to the other authors' application of codes, confirming their coding had been aligned with other coders. }

\new{The twelfth and final interview served as a check on convergent code definitions and saturation. Both co-first authors coded this final interview independently using the final code definitions for the axes of resonance, and independently found them to coherently apply to the twelfth interview as well, suggesting code saturation \cite{hennink2017code, buckley2022ten}. Furthermore, no new themes emerged from the twelfth interview for either of the two coders, suggesting thematic stability~\cite{saunders2018saturation}. The two co-first authors compared their independent application of existing codes with respect to the final interview, and agreed on the application of codes, with minor discrepancies in code application resolved through consensus discussion.}

\section{Results}
We found our participants used AI during the interpretive workflow in a variety of ways related to  supporting  them through challenges or uncertainties that arise,  providing \new{tools for perspective} that deepen their interpretations, and expanding their existing practice of divination, which we unpack in turn (see Figure 1, orange). \new{In this section, we first elaborate on our empirical findings, and then connect these findings to Resonance Theory along each axis.}

\subsection{Empirical findings}

\subsubsection{Characterizing AI in the broader tarot ecosystem}

\new{We first reflect on how AI is being conceptualized in the broader tarot community. Our participants noted widespread negativity around the use of AI in tarot (P1, P3, P4, P5, P9) but also that there is increased use of AI for tarot (P9). Our participants also highlighted concerns for its use, such as sycophantic properties \cite{sharma2023towards} (P3, P9), as well as concerns that incorporating AI into tarot readings will eventually result in deterioration of their intuitive skills (P3, P4, P5, P6, P9, P12).} 

\new{These widespread negative attitudes toward AI were also observed during our recruitment phase as well. 76.7\%\ of our requests to forum or server moderators to advertise our study never received a response, and 70\%\ of the responses we did get were rejections. Of the `no' responses, 57\%\ made explicit reference to our study's relation to AI in their reasoning to deny us recruiting in their community; some cited a forum-wide ban of discussions on AI, and others expressed concern that our study's subject would lead to arguments.  }

\subsubsection{Empowering users: navigating inner challenges or uncertainties that arise for the user}
One common motivation for using AI was to empower users to navigate moments of self-doubt or uncertainty that arose during the interpretive process. In these cases, AI functioned as a supportive presence by activating users' intuition, offering reassurance, and helping them work through ambiguous or emotionally difficult readings. 

Participants described turning to AI when faced with readings that felt especially opaque, involved interpersonal conflict, or allowed for multiple conflicting interpretations (P1, P2, P5, P6, P8, P9, \new{P12}). These situations often triggered uncertainty about their understanding of the cards or whether they could trust their initial instincts for an interpretation.
% the use of AI was to empower users by helping support and navigate challenges or uncertainties that arise for the user by supporting their intuition and offering assurances.  When dealing with ambiguous contexts (e.g. in interpreting a reading surrounding a particularly opaque subject, in navigating an interpersonal conflict, or where multiple contradictory meanings may exist), participants (P1, P2, P5, P6, P8, P9) used AI to mitigate self-doubt. Self-doubt manifested when participants felt insecure in their understanding of tarot card meanings, or unsure that they could trust their first instincts for an interpretation. 
Participant 1, for example, who has been developing his own divination systems for decades, described how by asking AI feedback on the details, he gained clarity and trust in his process:
\begin{quote}
\textit{``I've been working on the system for, you know, 30 something years, and I just second guess myself. Am I---am I overthinking this? Am I not thinking enough about this? Am I--- am I putting my biases on this too much? ... And in the last year, I would say, I really have kind of glommed onto AI as being very useful in helping me sort through the ideas about that. And I don't believe that AI has all the answers, but it is really a lot in as far as like, “oh, that's a good point”. “Yes, it kind of combines like that”, or “no, maybe not like that.''}
\end{quote}
% but has been able to quash these questions by asking AI for feedback on the details while he continues to refine this divination system. 
Participant 2 similarly shared that ChatGPT was a helpful tool in sorting out the self-doubt they often encountered during readings: 
\begin{quote}
\textit{``I have a tendency to not trust myself, or I have a tendency to let my fears kind of get in the way. And so with ChatGPT, specifically, I do have to ask it-- because ChatGPT can be a little bit of a yes man, you know-- but specifically asking for like, okay, where could my blind spots be in this situation, if X and Y happened? Or something like that, you know. What could this mean in the context of the reading? So I think, ...  typically, either like clarity, just to get rid of the muddiness or you know, if it's [a question/topic] that means a lot and I feel like, maybe I'm missing something.''} 
\end{quote}
When participants struggled to understand the context of card(s) in a reading, or how they connect to one another or to the question, participants asked AI for clarification on card meanings or for its interpretation of a spread (P2, P4, P7, P8).  Beginners also used AI as a learning resource, drawing on it for advice on interpretation, definitions of cards and symbols, and more, paralleling the role of books, tutorials, and mentors (P5, P7). 

\remove{{Connecting to Resonance Theory along the internal axis} With this pattern of results, we saw how divination is an interpretive labor that people find meaningful. The randomness of divination allows for rich internal spaces that are otherwise jammed by deterministic, mute relations. Internal resonance is about getting `in tune' with the self; it strengthens and validates the alterity, rather than suppressing it (via mute relations that are fostered through overthinking, or pressures). In the Jungian sense, it facilitates ``shadow work,'' shifting interpretative perspectives away from ego narratives and controlling outcomes and towards internal intuition. Participants found AI could empower them in fostering their internal intuition, which sets a design agenda for active engagement with interpretative labor. }

\subsubsection{Tools for perspective}
Another core motivation for using AI is its ability to act as \new{tools for perspective}, a way of deepening their interpretations by transcending their own subjectivity. In particular, participants drew on AI to refine, alter, and navigate a wide variety of interpretive lenses for constructing their interpretations. A widespread example of this, discussed by nearly all participants, was AI’s capacity to provide an alternative perspective. P5, as an example, shared that she often finds herself \textit{``focusing on a layer of the issue [whereas] the AI is focusing on a different layer. And what actually ends up happening [in future-oriented readings] is the AI layer"}. Receiving interpretations from AI allows her to discern which layers of a given problem she should hone in on, allowing her to refine her perspective. P8, who noted he involves AI in his readings ``not as a deciding party, but simply as an opinion", weighted AI slightly less in his meaning-making process but still valued it as a complementary viewpoint.

Participants also used AI to identify blind spots in their own interpretations and reveal details they overlooked in cases of ambiguous meaning (P1, P2, P3). As P2 described, using ChatGPT helped them \textit{``see through my own blindfold from like a different perspective.''} Similarly, P3 described a method she developed to fill in any potential gaps in her interpretations, which primarily involved asking AI for multiple alternative perspectives on a reading: 
\begin{quote}
\textit{\new{``I will often ask ChatGPT to give me 3 different interpretations. And I will reflect on each one with sort of an open mind and open heart, and think, is this possible that potentially one of these interpretations that isn't the one I would go to right away actually fits the situation....}}
    \textit{\remove{``And then the second thing I'm thinking, is it that, you know, maybe this does fit the situation, but I there's a blind spot on my end that I'm not seeing. So some things I might do. I might ask ChatGPT to show me the Python code to make sure that it actually pulled the cards at random.} I might ask ChatGPT for 3 alternative interpretations, and look at the other ones and see if one fits better. So that [one] interpretation it gave me maybe was fitting for narrative coherence, but there are other interpretations that resonate more and make more sense... none of them would have been interpretations I would have jumped to in my mind, but in retrospect, based on the commonly accepted card meanings, one of them maybe makes a lot of sense and reveals some insight that I wouldn't have thought of unprompted.''}
\end{quote}

Beyond offering alternative perspectives, participants also described AI as having the capacity to provide an ``objective'' account (P1, P2, P3, P4, P6). While views on how objective it truly was varied across participants, they viewed AI as a third party which does not entirely share their personal biases and has ``no stake'' in the topic. P2 argued that \textit{``it's difficult, I think, ultimately, to see things objectively for yourself...and so that's when I'll reach out to ChatGPT''} in order to use its outsider perspective as a point of reference when dealing with questions where they might not see themselves as objective. In contrast, some participants expressed doubt in AI's objectivity, such as P9 who stated that they know AI \textit{``is designed to tell me exactly what I want to hear, so I don't know what its process is when it [gives a tarot reading]. So I'm just assuming that it might not be as random as I would think it is.''} Yet even with this skepticism, they still used it as an outside perspective. 

Ultimately, whether or not participants regarded AI as objective, they found it a useful interpretive resource. As P8 described, AI represents \textit{``the culmination of human knowledge so far in the world.''}\remove{ representing a \emph{totalized Other}.} Because they saw AI's point of view as distinct from their own, participants across the sample drew on it to generate alternative interpretations of spreads in relation to their queries (P1, P2, P3, P4, P5, P6, P7, P8, P9, P11, \new{P12}). 

\remove{{Connecting to Resonance Theory along the horizontal axis}
With this pattern of results, we saw how AI was used to foster a form of relating that is not apathetic, alienated and indifferent. People are drawn to understanding and experiencing the perspectives of others, and engaging in social connection. Our participants used AI systems to encounter other subjectivities, as well as their own, casting new layers of sociality onto a traditionally solitary activity. }

\subsubsection{Streamlining and expanding the practice of divination}
One motivation for using AI was to efficiently streamline and expand the existing practice of divination across the entire tarot workflow, from i) pulling cards to ii) interpreting those cards to iii) writing down results. 

\paragraph{\textbf{Pulling cards and devising randomness}} Some users used AI to virtually draw cards, as opposed to using a physical deck in tandem with AI as this was faster than shuffling a physical deck multiple times and drawing the cards yourself (P3, P6, P9, \new{P12}). In addition to simulating card draws, one participant found other ways to generate randomness from the AI. Participant 9 said that hallucinations in LLMs could be used for randomness:

\begin{quote}
\textit{``If AI has a hallucination while I'm doing tarot, that is almost that could almost be part of the reading, because in that case you kind of see the thing that stands out. That isn't true, as I don't know. That's one of those things to pay attention to, just like a card falling out of the deck, or something.''}
\end{quote}

They also emphasized that the visual indeterminacy of generated imagery was a source of randomness useful for divination. Recalling their use of Midjourney to create images as objects of interpretation, they explained:
\begin{quote}
\textit{``I had it generate an image of a tarot card---just a tarot card, didn't tell it which one--and so it kind of spat out this sort of amalgamation, of, of what it thought a tarot card should look like. But I thought, that's an interesting method of divination, it's almost like a Rorschach test, where …  it tells me what it thinks. You know, what card it's thinking of. So I'm going to see certain symbols just randomly pop up there. There's sort of a method of that level of randomness there.''}
\end{quote}

\paragraph{\textbf{Reading tarot cards}}
Nearly all participants directly asked AI to do the reading itself; which involved having the AI generate and write out interpretations. One common pattern of use was interpreting the results of a tarot reading themselves first, before involving AI in any capacity. P7, who predominantly used web-based tarot reading platforms which generated AI interpretations for the user, stated \textit{``I start seeing the readings and if it makes sense or not, or if it could provide me some guidance. So first, if I see the explanation from the AI, I first won't [read] it. I first want to [interpret the cards] myself, and then secondly... I start to compare it with the AI’s,''} using a process of first interpreting a reading himself and then assessing the reading holistically by also taking the AI-generated interpretation into account.  \remove{Two} \new{Three} participants even let AI do all of the interpretive work (P4, P10, \new{P12}), only evaluating the tarot spread themselves after reading the interpretation provided by AI. P10 described his personal process, where he allows AI to do the brunt of the interpretive labor:
\begin{quote}
\textit{``I'll just pick the card, go to...my AI... I give it a prompt... These are the cards that [I pulled]. And without bias, it should give me… what is the reason of what each of these cards means, so it should make a cohesive response, and a very sound judgment for me.''}
\end{quote}
P4 likened the process of understanding the meanings of multiple cards within a spread to \textit{``doing sudoku in [her] head,''} and found that using AI to interpret a reading for her greatly reduced the energy required for interpretation. \new{P12 described a similar experience, finding that using AI for interpretation allowed her to engage with her process of setting intentions for the month ahead even when she had limited time available:} 

\begin{quote}
\new{\textit{``It's usually going to take me at least, like... 30 minutes to 1 hour to be able to figure out the meanings, to be able to also set up the intentions, try to connect everything, and there's so many informations usually popping up in my mind, too, so it's really… If I'm really stressed, then [doing a reading] is going to be kind of more stressful for me as well. And for the AI ones, usually I'm just using that to set up a quick intention for the future, and also these days, I'm using that to set up some monthly tasks to experiment on as well, so… Just kind of an easier process for me, so to speak, on that end.''}}
\end{quote}

For \remove{both} \new{these} participants, offloading the majority of the interpretive process to AI allowed them to shift their focus away from determining the cards' meanings, and towards understanding and applying the results of the reading to their life. See Section 4.2 for further discussion of the AI tools people used in the interpretative process itself. 

% \subsubsection{Extending existing tarot practices with AI}

%  Many participants mirrored their tarot habits and practices when doing readings assisted by AI; [insert how doing this streamlines and/or expands their divination practice]. [textual support goes here]
% In the words of Participant 6: ``while... the algorithm is trying to process everything. I focus a lot on the question. So let's say, for instance, I'm asking, what will the outcome be if I go to this place? And when I tap that in, I focus on the question while the algorithm is processing.''

\paragraph{\textbf{For documenting a tarot reading or looking up references}}

Participants used AI to journal tarot readings through writing out interpretation and quickly jotting notes about a reading (P4), storytelling (P1, P9), and for feedback on their own writing (P1, P9). Users also used AI as a search engine to source information both related to tarot and divination more broadly (P4, P5, P7). In particular, this manifested as looking up card meanings and the established image symbolism (P4). Two participants used AI to put into practice ideas or concepts they had for a new divination system, and AI provided feedback to solidify the system (P1, P9). Across this variety of AI use cases, we found that more experienced tarot practitioners were generally more critical about the use of AI. Advanced participants worried that becoming too reliant on AI would eventually cause their intuitive skills to deteriorate, which is a sentiment reflected in this comment from P9:
\begin{quote}
\textit{``It could be really easy just to go ask [AI], you know, what the answer is, and if you do that you're not going to get the ... subconscious moment where you're intuiting the experience yourself. Which is why I think it's really important, you know, when I do use it, that I make the interpretation myself before I go ask for it.''}
\end{quote}
Both intermediate and advanced participants expressed concerns about generative AI's impacts on the art industry, especially those who had their own art practices on top of tarot (P2, P5, P6, P9). Participants who were newer to tarot, however, were more unequivocally positive about both their personal use and the broader implications of AI, and did not share the concerns of the more advanced practitioners about AI inhibiting intuition (P4, P7, P8, P10, P12).

\subsection{Connecting empirical findings to Resonance Theory}
We examine how tarot activates the axes of Resonance Theory and how AI shapes meaning-making along those axes.
\subsubsection{Along the internal axis} Almost all participants (P2, P3, P5, P6, P7, P8, P9, P10, P11) report that tarot provides a way to strengthen connection to the subconscious or intuition---a sentiment that aligns with Rosa's internal axis of resonance. Participants described the labor of interpretation as a practice of getting in tune with their subconscious, using techniques like journaling (P2, P9) or mindfulness (P2, P8, P9, P10, P11) to deepen the connection and improve their readings.

However, many participants found that the instantaneous answers provided by AI inhibited their intuitive grasp of the card's meaning and thus negated this sense of self-resonance (P3, P4, P5, P6, P9, P12). Participants (P8, P9) often disciplined themselves to first making interpretations without AI:
\begin{quote} \textit{``It could be really easy just to go ask [the AI]. You know what the answer is, and if you do that you're not going to get the experience. You're not going to have the subconscious moment where you're intuiting the experience yourself. Which is why I think it's really important, you know, when I do use it, that I make the interpretation myself before I go ask for it.''} (P9)
\end{quote}

However, some participants (P2, P10, P11) were able to exploit specific features of AI to empower their self connection. For example, P11, who saw themselves as an AI optimist relative to their friends, found the dialectical nature of ChatGPT could ``deepen the reading'' when prompted to suggest new journal prompts or ways to explore the themes practically in their life.

\subsubsection{Along the horizontal axis}
We saw that tarot users are interested in understanding and experiencing the perspectives of others, evoking a connection with Rosa's Horizontal Resonance. 
Three distinct forms of sociality emerge from this data. First, almost all participants reported a strong desire to reflect on interpersonal relationships, and ask questions related to friends, families, and romantic partners not present (P1, P2, P4, P5, P7, P8, P9, P10, P11, P12).

Another form of sociality reported was the sociality of the tarot reading itself, connections made with other people present during readings (P2, P5, P6, P7, P11), invited in to receive readings (P1, P2, P3, P5, P6, P7, P8, P9, P11, P12), as collaborators providing secondary opinions (P2, P3, P5, P6, P7, P8), or as teachers of the practice (P2, P4, P6, P7, P10, P11). This ``human factor'' (P8) was found to be a repeated theme, allowing readers the pleasure of ``seeing other worlds'' (P9). Participants brought in AI systems for a similar purpose, as a way to encounter other subjectivities and perspectives during solitary readings.

At its most extreme, a final form of sociality identified was a parasocial connection between the readers and the AI system itself (P1, P2, P7, P8, P9, P10, P11). %Participant 2 even give their AI chatbot a name as a way to honor their repeated relationship with it. WWhile participants clearly value their interactions with an artificial other, as a consequence of instantaneous access to neutral alternative perspectives, 
Many participants (P2, P5, P7, P8, P9, P10) report seeking less interpretative advice from their human friends, in lieu of turning to AI. Many highlighted that convenience of AI gave them increased access to alternative perspectives, particularly in cases where they did not want to ``bug'' (P8) or ``pester'' (P9) their human friends: \textit{``But sometimes it's nice to have that extra perspective, and my friends aren't always around, and also my friends know me, and they'll like be nicer. So it's nice sometimes to just have like an objective perspective.'' (P2) }
%Given that the interpretive power of tarot  rides in experiencing horizontal resonance, this suggests a need for design recommendations that acknowledge the potential parasocial relationships being made between humans and AIs and how they impact human-human relationships.} 

\subsubsection{Along the diagonal axis} Encounters with the material affordances and ritualistic practices of tarot are foundational to the resonance experienced by readers. Most commonly participants report rituals to promote randomness (e.g., shuffling the cards) (P1, P2, P3, P7, P8, P9), rituals surrounding the numeric and spatial arrangements of the cards (e.g., pulling three or five cards at a time) (P1, P2, P5, P6, P9), rituals to induce mindful (e.g., deep breathing) (P2, P3, P5, P8, P10, P11), or rituals to mark a transition to a mindful or spiritual mood (e.g., lighting incense) (P1, P2, P3, P11). Additionally, the card imagery, beyond the known conceptual meanings, acted like scaffolding for readers, who could make inferences from the visual affordances of the cards in moments of uncertainty or doubt (P1, P5, P7, P11, P12).

%This form of resonance found through artifacts and activities links to the pre-modern (or non-modern) worlds of animism, of things having life, of the tarot deck as a living thing that needs to be tapped to awaken it, a sentiment discussed by P2: }
%\new{
%\begin{quote}
%``I'll like spread the cards out...Fan them out, and then I'll kind of run my hands gently over them, and see which ones kind of like follow me. When you're shuffling cards that jump or just like, you know, just fall out. Or you know typically what I do. When I pull a card for oracle cards I'll shuffle, wait for them to jump.'' 
%\end{quote}}

Many readers experimented with ways AI-based readings could extend or alter these material rituals, for example by inducing more card draws as a consequence of increasing the ``margin of error'' (P8), ensuring randomness through a random number generator (P3), or inducing AI hallucinations (P9). However most recognized that as ``machines'', these systems operate on ``algorithms instead of mysteries'' (P7). \begin{quote}
    \textit{"It's a different feeling. It's a different energy. It's different technique. Different meaning. Because I think with chatGPT, you're kind of pulling more from all of the the words that you pulled that you put into it over time. So it's kind of drawing from more of a cumulative energy that you've sort of left behind."} (P9)
\end{quote}

%This sets a design agenda for AI systems to support diagonal resonance via injected serendipity that aligns with the unique affordances of machine learning, rather than the existing affordances of tarot experiences. \cite{ugander2024art}.}

\subsubsection{Along the vertical axis}
For many readers (P2, P3, P6, P7, P9, P10, P11) reading tarot gave them a connection to higher unknown forces and their broader spirituality. The ``divine forces'' (P2) and unknowable ``mysteries'' (P7) are evoked by the specific practices of tarot:

\begin{quote}
\textit{``it's that intuitive nudge, either from your guides, or forces that are protecting you, guiding you, and also  your own divine energy in yourself guiding you towards certain cards, when you're shuffling cards that jump or just fall out.''} (P2)
\end{quote}

This connection akin to Rosa's vertical resonance seemed highly personally motivated, related to the religious backgrounds of the readers (P9, P10). Only one participant spoke to the potential for AI to \textit{``ruin the mysteries''} (P7) by making the process appear understandable.
Some participants (P2, P6, P11) reported extending their use of AI to spiritual practices beyond tarot, like channeling spirits:
\begin{quote}
    \textit{``Basically, there's a process under which we try to connect the spirit to a certain object, and the theory that we've been working on is connecting the spirit to something like an AI. So by that process it could be possible to connect the spirit to something like chatGPT or DeepSeek.''} (P6)
\end{quote}

Given the rise of spirituality in technological discourse, this axis of resonance and its relationship to AI deserves deeper investigation in future work.

\section{Discussion}

%What is special about tarot? Why do we care? 
Unlike self-dialogue, tarot divination mitigates the risk of closure and solipsism. Self-dialogue can easily fall into repetitive loops, rehearsing the same thoughts without opening new directions. By contrast, the symbolic and open-ended nature of the tarot cards cultivates receptivity to meanings and perspectives that might not surface through internal reflection alone. Random card selection adds contingency to the process, introducing unexpected prompts that can shift the direction of thought. In this sense, tarot provides a structured form of alterity---an encounter with something external that resists complete control or closure. The frame imposed by a randomly drawn card creates interpretive possibilities that self reflection, by itself, cannot easily generate~\cite{holquist2003dialogism, aphek1990semiotics}.

% Unlike self-dialogue, tarot divination mitigates the risk of closure and solipsism, with the symbolic, open-ended nature of cards cultivating receptivity to new meaning and perspectives. The randomness inherent in card selection introduces contingency into the interpretive process, creating space for unexpected perspectives. Furthermore, the alterity often missing from internal dialogue is enriched by the external symbolic structure that tarot provides \cite{holquist2003dialogism}. Tarot enables dialogic engagement beyond the self \cite{aphek1990semiotics}, by introducing encounters with otherness. The frame introduced by a randomly pulled tarot card opens interpretive possibilities unavailable to purely internal reflection. 

Resonance Theory gives us a framework to articulate and decompose these interpretive possibilities. Further, it offers a blueprint for thinking through the role of AI in the process of meaning-making. Our participants report that encounters with AI interpretations opened useful interpretive possibilities. We repeatedly heard how users were able to use AI to see the same data through another lens. This is potentially very powerful. However, AI can fulfill many roles, only some of which actually enrich this interpretive possibility. AI chatbots are known to be sycophantic \cite{sharma2023towards}, confirming what a user already knows or the interpretation they already have. At another extreme, AI may also be overly authoritative, communicating with certainty about an interpretation that amplifies a user's existing tendency to ``anchor'' on whatever interpretation is given first. Participants encounter both extremes and were often critical of AI interpretations. Nevertheless, participants were often able to use AI to help them see the data through another lens, helpfully introduced from outside themselves. As such, visions of AI in interpretive positions that offer fluid alternative interpretations can be useful. We believe it can leverage the same power of randomness that tarot does: the opportunity to use a randomly generated symbol to great symbolic effect. This motivates and grounds a series of design recommendations for designing AI systems for interpretative tasks.

\subsection{Design Recommendations}
This next section provides insights into how these insights create new potentials for the design of sociotechnical systems which scaffold the meaning-making process.

\subsubsection{Strengthening the ``intuitive muscle'' through active engagement}
Evident from these interviews, interpreting tarot symbolism is an actively engaging process that creates value for users asking big life questions. While these AI-enabled readers could have completely relied on AI to provide them with concrete, specific answers, few chose to leave the task of interpretation solely to the model. This is true across all experience levels with tarot. While the integration of chat-bots into their practice differs, including the degree to which the AI is doing labor, most readers first attempt to make their own interpretation before consulting the AI. This \remove{consistent }behavior speaks to a common value around maintaining some degree of agency in interpretation, and seems to emerge because these users value the process, rather than just the product of divination. In other words, there is a desire to do some level of interpretive labor, because practitioners enjoy the process of gaining expertise, and cannot achieve a meaningful reading without doing some kind of work. Therefore, AI was the most helpful when participants used it to navigate challenges or uncertainties by strengthening of their own intuition through active engagement and in turn, fostering internal resonance. 

It is important to note that the scope of this research is limited to tarot readers, not tarot consumers. In the long history of divination, most users of divinatory systems have no experience reading symbols and so heavily rely on spiritual leaders to conduct and read the symbols for them. In these contexts, the seeker relies on the practitioner. It is unclear if a wider consumer-oriented audience, which seeks questions from AI would treat their interactions with AI as an opportunity to strengthen their intuition for tarot symbolism, or use it as a rapid Q\&A.

Nonetheless, given the goal that many HCI designers hold, to develop systems which do not lead to problematic reliance on AI systems, the case of tarot card readers and chat bots is a unique guiding example: the context of tarot creates strong motivations to make the process of strengthening one's intuition enjoyable and necessary for the desired end-state to be achieved. Thus, designers creating experiences for meaning-making could consider how the overall context of the experience scaffolds a unique skill as a necessity for understanding: the skill of meaning-making. Since we saw that deeper insights, and more understanding can be achieved through practice and self work, we believe the addition of ambiguity, providing limited responses \cite{kalai2025hallucinate}, or adding learning objectives, could be employed to accomplish this goal. Furthermore, the sycophancy that RLHF induces \cite{sharma2023towards} may be at odds with the project of meaning-making, a tension many of our participants had to navigate.

Additionally, this work calls for design for multiple modes of self and other reliance. In some cases readers wanted more help from the AI, and in others they wanted to focus mostly on their interpretation. A sliding scale which enables a user to flow through multiple levels of their own self-declared ability could be an important feature of such interfaces. If intuition is a muscle, then designers ought to consider patterns that ``work'' that muscle, and prevent the creation of over-reliance. 

\subsubsection{Friction as a force for meaning}

In the context of AI-assisted tarot, the unique affordances of chat-bot interactions, gave rise to new forms of divinatory practice. By appropriating the dialectal nature of chat-bots, participants were able to receive alternative perspectives on the meaning of cards. These experiences allow readers to pause, reflect, and challenge their thinking beyond interactions with a static guidebook, with most readers actively seeking out alternative responses by requesting multiple interpretations, and trying out different ``layers'' in the interpretive space (i.e., shifting contexts). If these alternate interpretations aligned with their current hypothesis, the reader felt validated or reassured in their doubts about the interpretation. But more interestingly, the value of a dialectal partner was also in its ability to challenge their thinking. In cases where the AI disagreed with their interpretation, the reader was given an opportunity to deepen, clarify, adjust, or shift their thinking around a subject. Alternative subjectivities were seen as valuable partially because of their ability to increase friction and generate lateral ideas. 

Designers should consider how the value of co-interpreting meaning with an AI model lies beyond its ability to confirm or conform to a users perspective. The friction of encountering alternate perspectives gave rise to \textit{critical engagement} \cite{kawakami2023training, everson2022key}, where users could strengthen their own perspectives, rather than doubt them. One way to do so, which our participants actively explored, is to foster interactions with a range of alternative perspectives that subvert both dominant, hegemonic representations, as well as those of the individual themselves. Designers could intentionally design experiences which provide less perfect or incomplete responses, forcing the user to address the inconsistencies from multiple angles. Using multiple responses from multiple perspectives which expose the user to the wide range of potential solutions or lines of questioning is a strategy that might induce a kind of productive friction \cite{haghighi2025ontologies, park2025ask}.

We note that many readers in this sample took advantage of the perspective power of AI, and treated the interaction more like an interactive guidebook, rather than an active partner in providing alternate subjectivities. If designers wish to encourage unique reflective practice that act differently from search engines, they might consider making the affordances of interpretation more visible, and intentionally designing experiences which feel unique from current meaning-making practices. For instance: the AI could provide proactive suggestions, draw on unexpected contexts, or ask more probing questions. 

\subsubsection{Re-injecting randomness into generative AI systems}
In our participants we found \remove{great care with the rituals of randomness}\new{intentional interactions with randomness} that helped give meaning to the process of tarot. Indeed, the randomness inherent in tarot introduces contingency into the interpretive process that creates space for unexpected perspectives and serendipitous encounters. This enables dialogic engagement beyond the self by opening up a rich variety of possible perspectives \cite{aphek1990semiotics}. Our participants intuited this, and their uses of AI in their tarot practice navigated and re-imagined randomness in creative ways: many readers did use AI to do card pulls, even when they were using AI as a disjoint experience separate from non-AI tarot practice, because they valued the act of generating randomness itself, seeing it as central to what gives divination its meaning. When others did seek randomness within AI systems, they found creative ways to do so, such as producing AI-generated imagery or asking ChatGPT to use Python's \texttt{random()} in order to draw tarot cards for them. 

These encounters point to an alternative way of interacting with AI that considers randomness a feature, not a bug---a design principle that can foster serendipity and alterity in its capacity to promote active engagement and ritualized meaning-making. For example, in their critical art-making practice of AI divination, \citet{schroeder2025ai} find that too much technological determinism undermines the expressiveness of their system, and use a physical bingo cage to \remove{set}\new{anchor the} random seeds of their generative model \new{to physical randomization rituals}. Following from this principle, designers could leverage meaning made from randomness by embedding the delivery of data, results, or information within a limit context resembling a ritual. For example, designing thoughtful experiences around what information from a large dataset is presented first, or drawing on meaningful physical or embodied symbols (e.g., moon cycles, seasons) to influence random interface events \cite{isbister2019toward}. 

\section{Limitations and Future Work}
\new{As an exploratory probe to develop concepts in this understudied domain, we note that} our work has several limitations which motivate future work.  First is our small sample size of only \new{12}\remove{11} participants.  As discussed above, the use of generative AI within the tarot community is highly polarizing, which limited our capacity to recruit participants through traditional methods. Although these topics have international influence, we were limited to a sample of North American participants due to IRB restrictions and these limitations with participant recruitment. \new{We acknowledge that size and selection effects limit the generalizability of these findings.
} Future work should expand \remove{the number of participants }beyond the North American context \new{to participants from other countries and non-English-speaking participants and correspondingly} identify means of recruitment that engage a broader selection of tarot users who use AI. \new{In particular, due to the stigma around both tarot reading itself and the use of AI for tarot reading, future work should explore anonymous methods for interviewing.}

We note that because of these selection criteria, tarot practitioners who have interacted with AI in their practice, they are inherently a cohort with some interest and appreciation towards AI interaction. This stands in contrast to the tarot community at large, which largely harbors negative sentiments toward AI. Future work is therefore required to understand the nuances of the broader tarot communities' attitudes toward AI and how they fit into our theoretical account. 
It is also important to keep in mind that these results are critically shaped by the level of participants' expertise in tarot practice and the tarot practitioners' trust in AI, which future work should more carefully understand and see how those key parameters interact with the findings discussed here.

\new{We also note that AI systems continue to evolve as they absorb new data and cultural material. As their outputs shift in response to these expanding corpora, the symbolic possibilities they generate will likewise diversify, enabling interpretations that become richer and increasingly layered. It also remains to be seen whether these emerging AI-mediated practices will feed back into the broader tarot community, particularly among readers who currently reject AI, and whether divergent interpretive norms will begin to form across groups.}

\new{Our work opens the door to several future research questions. Future work should explore how trust in and reliance on AI impacts reader's attitudes and interpretations. Furthermore, it should explore how the degree of tarot experience affects perceptions of AI's capacity to support tarot interpretations. Finally, beyond the esoteric use case analyzed here, future work could explore more mainstream uses of AI for meaning-making.}

\remove{We also note that from our participant interviews, we were able to identify and map three of the four axes of Rosa's Theory of Resonance. In particular, the vertical axis of transcendent and existential resonance did not emerge as a theme in our coding. P9 did hint at the connection between randomness and vertical resonance when they said that ``there's a consciousness in the universe. That pervades all things, and I feel like there's ebbs and flows to it. There's currents to it, and I think that tends to manifest. One way it tends to manifest is just randomness. Anytime you get just pure randomness. And so with the tarot, it's taking something totally randomly generated, and using that as kind of a conduit between me and that ebb and flow, and looking for patterns that are sort of in the current.'' This idea is in line with the beliefs of chaos magick---which creates arbitrary links and provisional correspondences as a central tenet of its cosmology. But ultimately future work is needed to explore how participant's theorize randomize and how that connects to their cosmology and uses of AI.}

% We also note that as AI systems are constantly evolving as they access new information. It is important to note that the symbolic effects of AI’s randomly generated outputs will continue to expand as new data and cultural material emerge, opening the door to increasingly layered and richer interpretations.

\section{Conclusion}
\remove{We have so far examined how tarot practitioners use AI to extend their interpretative capabilities while asking big life questions. We believe that these detailed accounts provide a unique lens to explore how AI models can and should be used ethically during experiences relating to meaning-making and life reflection. In this paper, we focused on AI-assisted tarot reading, a divinatory practice where users accompany their query with randomized data (i.e., drawn cards) and ask AI to interpret the data in response to their query.}\new{This paper offers the first empirical account of how tarot practitioners integrate AI into non-causal, interpretive meaning-making. Drawing on 12 interviews, our findings surface previously unarticulated dynamics in how practitioners preserve, share, and sometimes cede interpretive control---decisions that shape the authorship of meaning in AI-assisted divination.} \remove{Drawing on Rosa's sociological Theory of Resonance and qualitative interviews, we explored how users negotiate meaning with AI in a context where no causal link exists between the data and question. We find diverse uses of AI in the interpretive process of tarot reading, including helping participants navigate challenges or uncertainties, providing ``perspectivist powertools'' to navigate and explore a wide variety of perspectives and streamlining and expanding the practice of divination.}\new{We believe that these detailed accounts provide new knowledge about how AI models can and should be used ethically during experiences relating to meaning-making and life reflection.}
\new{Using Rosa's Theory of Resonance, we reveal how AI introduces forms of resonance that practitioners actively work with rather than passively receive. We argue that AI systems for interpretive tasks should foreground user agency by offering multiple perspectives and preserving openness rather than narrowing meaning. }
\remove{Through designing AI systems for interpretative tasks that should foster active engagement and serendipity through injected randomness and the friction of many perspectives, these tools  can offer an alterity that helps break us out of the shackles of our minds and find resonance with the broader world.} \new{We call on future work to consider how AI systems can bring people back to slower, more reflective forms of meaning-making, where randomness, ambiguity, and openness create the conditions for resonance rather than closure.}

\begin{acks}
We are grateful to our participants for their time and for openly sharing their experiences and practices.
\end{acks}

\section*{Positionality Statement}
We reflect on how our own positionality may have influenced this study. The authors’ research subjects range broadly across disciplines, including HCI, media studies, communication, and the history of tarot. We all have some level of connection with tarot and divination---whether as hobbyists, or on a much more personal level---and the author team combined has been practicing tarot for over \remove{10}\new{40} years. We all also have personal views on the ethics of applying AI to interpretive tasks and a subset of us have worked together in a spiritual art collective for the past 2 years in refining and sharing a concept of AI divination at the intersection of our shared artistic, academic, and spiritual practice. 

%In this work we aim to evaluate AI’s role in divination in order to provide design recommendations which further support the user in polysemic interpretive tasks.

\bibliographystyle{ACM-Reference-Format}
\bibliography{bibliography}

\appendix

\section{Research Methods}

\subsection{Screening Survey Questions}
\begin{enumerate}
    \item Please briefly describe your experience with a recent or memorable tarot reading. This can include the type of spread you did, the card(s) you pulled, and/or your response to the card(s).
\item Have you ever used AI for helping you interpret a tarot/oracle spread?
\item Please briefly describe your experience.
 What specifically interests you about participating in this study?
\end{enumerate}

\subsection{Interview Questions}
Personal Background \& Practices

\begin{enumerate}\item How experienced are you with tarot? Would you say you are a beginner, intermediate, or expert? How long have you been practicing tarot?
    \item What first drew you to tarot? How did you start learning to read cards?
     \item How does a tarot spread produce meaning, philosophically?
      \item What types of questions do you typically turn to tarot for? (e.g. emotional/spiritual guidance, divination, etc.)
       \item Can you describe what you usually do when you sit down to do a reading—for yourself or someone else?
        \item Do you typically use spreads? If so, how do you decide what spread to use? 
         \item When you see a card (in the context of a spread or in a single card pull), what’s your first step in figuring out what it means in that reading?
          \item Do you mostly read for yourself, for others, or both? Does your approach change depending on who the reading is for?
\end{enumerate}

Interpretation Process
\begin{enumerate}
\item In this section we are curious a bit about the interpretation process in general. To start, in the screening survey you shared: [insert screening survey Q]. Can you tell me a little bit more about that (experience, how the reading felt, the impacts, etc)?
\item Can you recall a time where, in a reading, the cards didn’t seem to paint a coherent picture? What kinds of things go through your mind as you figure out how the cards fit together? (if they can’t remember a specific instance) Can you tell me a bit about how you might address a spread if cards seemed confusing, or contradict one another?
\item  Do you ever look up card meanings—online, in books, or other sources? When you decide to do that, what are you hoping to get from it?
\item Once you’ve looked at all the cards, how do you put them together into one overall message or story?
\end{enumerate}

Use of AI Tools
\begin{enumerate}
\item You mentioned in the screening survey that [how they use AI for tarot]. Could you tell me more about that? 
\item How often would you say you use AI for tarot/your practice? (e.g. always, occasionally, rarely)
\item Do you ever intentionally avoid using AI for tarot/your divination practices? If so, when and why?
\item Does your tarot process differ when using AI from your standard practice? (E.g. if you focus on the question when doing a reading without AI, do you also focus on the question when doing a reading with AI?)
\item Are there any particular situations or questions that draw you to using AI? Or is it part of your default practice?
\item In your perspective, what types of benefits does AI afford? What are its drawbacks?
\item Do you find the AI responses to be more generic or specific?  If AI’s responses are too generic, how do you respond to make it provide answers that are more applicable to your question(s) or life circumstances?
\item Do you ever feel like the AI responses fall short? How do you respond to get the answers you want?
\item Does what the AI says ever affect how you ended up interpreting the spread? If so, how?
\item In your experience, would you say there are types of questions or spreads where AI gives more, or less, helpful answers?
\item To get a good response from the AI, do you feel like you have to explain your situation or provide background? How much did you need to share for it to become accurate?  Is there any information that you specifically choose not to share with AI? If so, why?
\item Have you ever felt like the AI’s interpretation was off? If so, did you ignore it, adjust it, or something else?
\item Have you ever felt like the AI’s interpretation didn’t make sense at first, but then later you saw it differently or came to agree with it? 
\item Have you asked other people to interpret a spread for you? If so, how do you think AI’s interpretations differ from that of people’s?
\item Does using AI change how a tarot reading feels to you, whether emotionally, spiritually, or otherwise? If so, how?
\item How have your practices changed due to the introduction of AI tools? 
\item How do you feel about AI outside of tarot/your divination or spiritual practices? Would you say that you’re more pro- or anti-AI overall?

\end{enumerate}

\end{document}